\documentclass[aps,prb,twocolumn]{revtex4-1}
\usepackage{hyperref}
\usepackage{graphicx}
\usepackage{amsmath,amssymb,amsfonts}

\begin{document}

\title{Crystal structure and magnetic properties of Ba$_2R_{2/3}$TeO$_6$ ($R$ = Y, La, Pr, Nd, Sm-Lu) double perovskites}
\author{Tai Kong} 
\author{Robert J. Cava}

\affiliation{Department of Chemistry, Princeton University, Princeton, NJ 08540, USA}

\begin{abstract}

A new series of cubic double perovskites Ba$_2R_{2/3}$TeO$_6$ ($R$ = Y, La, Pr, Nd, Sm-Lu) was synthesized via solid state reaction. The $R^{3+}$ and Te$^{6+}$ ions are ordered on alternating octahedral sites, with the rare earth sites 2/3 occupied to balance the charge. The lattice parameters decrease monotonically from a = 8.5533(3) $\AA$ for Ba$_2$La$_{2/3}$TeO$_6$ to a = 8.3310(4) $\AA$ for Ba$_2$Lu$_{2/3}$TeO$_6$. The lattice parameter for $R$ = Y is close to that of Ho. Analysis of the resulting bond lengths indicates a structural relaxation around the $R$ ion site. Ba$_2$La$_{2/3}$TeO$_6$, Ba$_2$Y$_{2/3}$TeO$_6$ and Ba$_2$Lu$_{2/3}$TeO$_6$ show primarily temperature-independent magnetic susceptibility due to the lack of a local rare earth moment. For Ba$_2$Sm$_{2/3}$TeO$_6$ and Ba$_2$Eu$_{2/3}$TeO$_6$, the susceptibilities are influenced by Van Vleck-like contributions from excited state multiplets. For the remaining members, the Curie-Weiss law is followed with low-temperature deviations that can be associated with various degrees of crystalline electric field splitting. No magnetic ordering was observed down to 1.8 K in any of the compounds.

\end{abstract}
\maketitle

\section{Introduction}

Perovskite oxides have been widely studied for the variety of compositions, structures, and properties that they display. They can generally be described as ABO$_3$ compounds where A is a large metal ion located in a cavity, often 12-coordinated, bordered by BO$_6$ octahedra sharing corner oxygens. Via chemical substitution, the perovskite family can be dramatically extended, for example, to double perovskite structures with formulas of A$_2$BB'O$_6$, AA'B$_2$O$_6$ or AA'BB'O$_6$ where the metal sites are occupied by more than one type of element\cite{King10}. Many different combinations of A, A' and B and B' are allowed, and thus in addition to the resulting structural complexities, perovskites offer many interesting physical properties, such as high temperature superconductivity\cite{Bednorz1986, Cava00}, large magnetoresistance\cite{Kobayashi98} and ferroelectricity\cite{vonHippel50}.

The first step in perovskite research is frequently to find and generally characterize new compounds with the perovskite structure. Although many double perovskites have been studied, relatively little is known about perovskites containing the Te$^{6+}$ ion\cite{Vasala2015}, whose size necessitates its B-site occupancy and whose charge requires at least a double perovskite to form for perovskite oxides. Recently, a new double perovskite, Ba$_2$Bi$_{2/3}$TeO$_6$, was found, with Bi$^{3+}$ partially occupying one of the two independent B sites\cite{Park00}. 

Formally, rare earth ions in oxides frequently have the same oxidation state as Bi$^{3+}$ with comparable sizes\cite{Shannon76} (e.g. $r_{Bi^{3+}}$ = 1.03 $\AA$, $r_{La^{3+}}$ = 1.03 $\AA$ and $r_{Lu^{3+}}$ = 0.86 $\AA$ for coordination number of six). Thus, in the current work we address the question of whether Bi$^{3+}$ can be replaced by trivalent rare earth ions in Ba$_2R_{2/3}$TeO$_6$ double perovskites, and whether insight can be gained about lattice stability in perovskites due to the fact that the rare earth radii can vary by as much as 30$\%$ across the series. Remarkably the double perovskite forms for rare earth ions from La to Lu. In addition to describing the structures of the new series, the magnetism of the rare earths in these double perovskites is characterized.

With this in mind, we synthesized 14 new cubic double perovskites, with chemical formulas Ba$_2R_{2/3}$TeO$_6$ ($R$ = Y, La, Pr, Nd, Sm-Lu). The crystal structures were characterized using powder x-ray diffraction. Our results, along with the previous reported Ba$_2$Bi$_{2/3}$TeO$_6$, show that the oxidation states as well as the relative sizes of the B ions play a substantial role in determining the stability of Ba$_2R_{2/3}$TeO$_6$ materials. The magnetic properties, systematically studied via temperature- and field-dependent magnetization measurements, are dominated by the rare earth magnetism, as expected, and show behavior consistent with observations in other rare earth oxide systems.

\section{Experimental Methods}
 
\textit{Synthesis}: Ba$_2R_{2/3}$TeO$_6$ compounds were synthesized from stoichiometric mixture of BaCO$_3$ (Alfa Aesar, 99$\%$), TeO$_2$ (Sigma-Aldrich, 99$\%$) and dried rare earth oxides (Y$_2$O$_3$, La$_2$O$_3$, Pr$_6$O$_11$, Nd$_2$O$_3$, Sm$_2$O$_3$, Eu$_2$O$_3$, Gd$_2$O$_3$, Tb$_4$O$_7$, Dy$_2$O$_3$, Ho$_2$O$_3$, Er$_2$O$_3$, Tm$_2$O$_3$, Yb$_2$O$_3$, Lu$_2$O$_3$ from Alfa Aesar, $>99.9\%$). The starting materials were ground and mixed in an agate mortar and then heated up to 1000 $^{\circ}$C in an alumina crucible in air. For $R$ = Ho-Lu, higher temperatures were used to obtain a pure phase. For $R$ = Ho, the sample was heated up to 1150 $^{\circ}$C and for $R$ = Er-Lu, 1300 $^{\circ}$C was needed. Pressing pellets and several intermediate re-grinding were performed to purify the phases. All final products are white and stable in air.

\textit{Structural determination}: Powder x-ray diffraction patterns were measured using a Bruker D8 Advance Eco, Cu K$_{\alpha}$ radiation ($\lambda$ = 1.5406 $\AA$), equipped with a LynxEye-XE detector. Rietveld refinements were performed using GSAS software\cite{GSAS,Toby2001}. The crystal structure drawing was plotted using the VESTA program\cite{Momma11}.

\textit{Magnetic property measurements}: Temperature-dependent (1.8-300 K) and field-dependent (0-90 kOe) magnetization measurements were made using a Quantum Design (QD) physical property measurement system (PPMS) Dynacool equipped with the vibrating sample magnetometer (VSM) option. Powder samples were loaded in standard QD transparent, plastic, capsules and mounted on a brass half-tube sample holder.

\section{Results and Discussions}

\begin{figure*}[ht!]
\includegraphics[scale = 0.58]{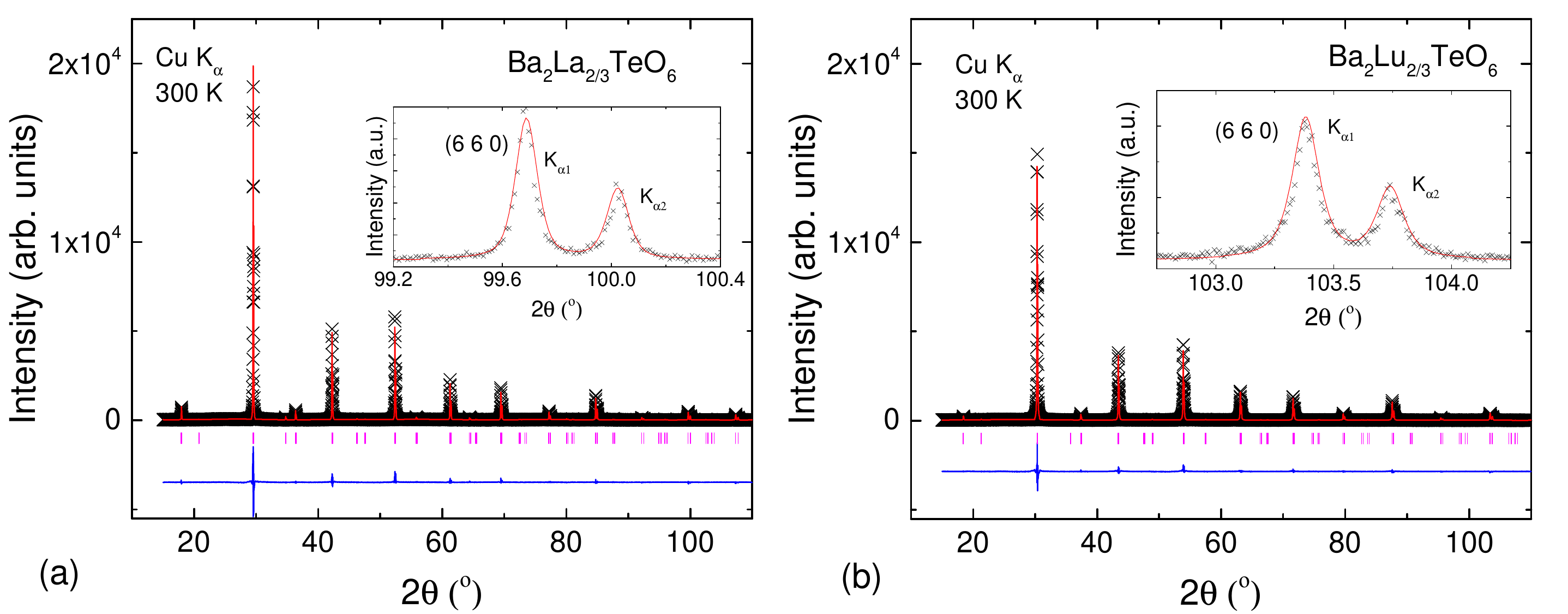}
\caption{(Color online) Room-temperature powder x-ray diffraction data for (a) Ba$_2$La$_{2/3}$TeO$_6$ and (b) Ba$_2$Lu$_{2/3}$TeO$_6$. Observed diffraction data are shown as black crosses. Calculated diffraction pattern are shown by red lines. The differences between observed and calculated data are shown in blue. Diffraction peak positions are indicated by purple ticks. Insets show (660) peaks at high angle.}
\label{LaLu}
\end{figure*}

Room-temperature powder x-ray diffraction data for Ba$_2$La$_{2/3}$TeO$_6$ and Ba$_2$Lu$_{2/3}$TeO$_6$ are shown in Fig.~\ref{LaLu} as examples. All observed peaks can be indexed with a cubic double perovskite structure. A simple perovskite structure, with the $R^{3+}$ and Te$^{6+}$ atoms mixed randomly on the B sites, aside from being highly unlikely due to the apocalyptic differences in charge and size between $R^{3+}$ and Te$^{6+}$ ions, is qualitatively eliminated by the observation in all samples of peaks that can only be indexed by a combination of odd indices on the double perovskite super cell (i.e. two times the basic perovskite cell of $\sim$4.2$\AA$) (Supplementary information Figure 1). In the insets of Fig.~\ref{LaLu}, the high angle (660) peaks are shown. Apart from expected double peak structure due to Cu K$_{\alpha 1}$/K$_{\alpha 2}$ splitting, no further peak splitting or asymmetry is observed, indicating a clear cubic structure for all the materials. Thus the Ba$_2R_{2/3}$TeO$_6$ double perovskite compounds crystallize in cubic symmetry, with space group $Fm\bar{3}m$ (S.G.225), with lattice parameters varying from 8.5533(3) $\AA$ to 8.3310(4) $\AA$ for Ba$_2$La$_{2/3}$TeO$_6$ to Ba$_2$Lu$_{2/3}$TeO$_6$.

\begin{table}[!h]
\caption{Refined crystal structure of Ba$_2$Gd$_{2/3}$TeO$_6$. Space group $Fm\bar{3}m$ (No. 225). Z = 4, a = 8.4253(3) $\AA$.}
\begin{tabular}{p{1cm} p{1cm} p{1cm} p{1.5cm} p{1cm} p{1cm} p{1cm} }
\hline
Atom & Site & Occ. & $x$ & $y$ & $z$ & 100U$_{iso}$  \\
\hline

Ba & 8c & 1 & 1/4 & 1/4 & 1/4 & 2.9(2)\\
Gd & 4a & 2/3 & 0 & 0 & 0 &  0.8(2)\\
Te & 4b & 1 & 1/2 & 1/2 & 1/2 & 0.8(3)\\
O & 24e & 1 & 0.267(2) & 0 & 0 & 1.1(5)\\
\hline
\end{tabular}
\label{structure}
\end{table}

\begin{figure}[tbh!]
\includegraphics[scale = 0.32]{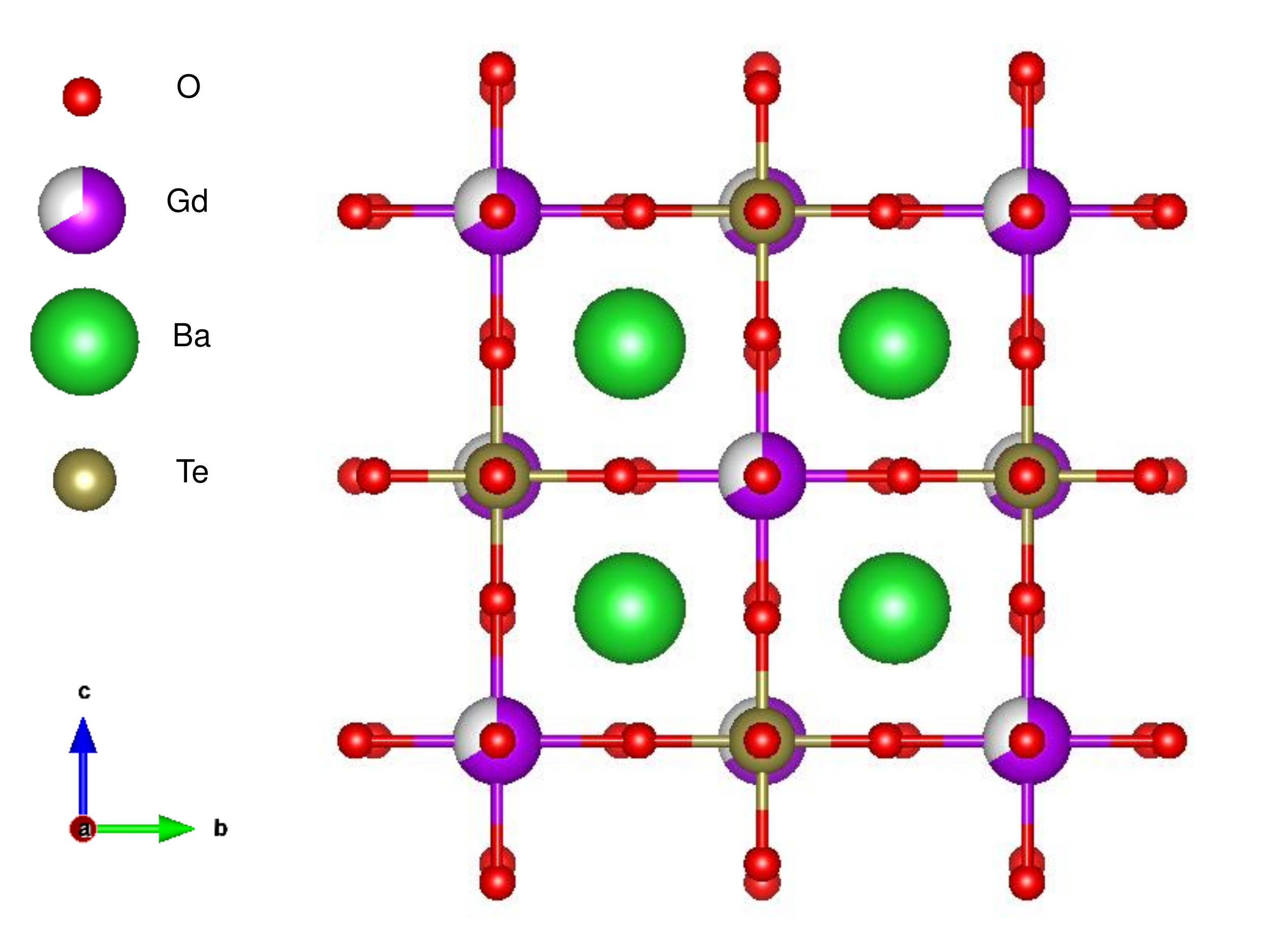}
\caption{(Color online) Crystal structure of Ba$_2$Gd$_{2/3}$TeO$_6$, with a-axis pointing out of the page. Oxygen is in red; gadolinium is purple with partial occupancy; barium is in green and tellurium is in brown.}
\label{strc}
\end{figure}

We employed the structure of Ba$_2$Bi$_{2/3}$TeO$_6$\cite{Park00} as the starting model for the Rietveld structural refinements. It is clear that our Rietveld refinement results, as shown by red lines in Fig.~\ref{LaLu}, mathc very well with the black observed diffraction data. The refined $\chi^2$ and $R_p$ values are 2.4 and 12.6$\%$ for La; 2.1 and 9.9$\%$ for Lu, for example, with only 1 positional parameter involved-the x parameter of the oxygen in the 24e sites. Although the precision of the x parameter determined by laboratory powder x-ray diffraction is only $\sim$1$\%$, it is sufficient for the present purposes. To further understand the crystal structure of Ba$_2R_{2/3}$TeO$_6$, we take Ba$_2$Gd$_{2/3}$TeO$_6$ as an example. Detailed crystallographic information of  Ba$_2$Gd$_{2/3}$TeO$_6$ are listed in Table~\ref{structure}. The crystal structure drawing of Ba$_2$Gd$_{2/3}$TeO$_6$ is shown in Fig.~\ref{strc}. In this double perovskite, the Gd$^{3+}$ and Te$^{6+}$ ions occupy the centers of GdO$_6$ and TeO$_6$ octahedra,respectively, which are then structurally arranged in the A$_2$BB'O$_6$ order found in the Elpasolite K$_2$NaAlF$_6$ structure type\cite{Moras74}. This B site ordering is consistent with the expected structure stability phase diagram\cite{Anderson93}. We can also calculate the Goldschmidt tolerance factor as\cite{Goldschmidt26}: $r = \frac{r_A+r_O}{\sqrt{2}(r_B+r_O)}$, where $r_A$, $r_B$ and $r_O$ represent the ionic radii of A, B and oxygen ions. In the case of Ba$_2R_{2/3}$TeO$_6$, $r_B$ can be approximated as $r_B = \frac{1}{2}(r_{Te^{6+}}+\frac{2}{3}r_{R^{3+}})$. The estimated tolerance factors for Ba$_2R_{2/3}$TeO$_6$ range from 1.05 in Ba$_2$La$_{2/3}$TeO$_6$ to 1.08 in Ba$_2$Lu$_{2/3}$TeO$_6$. These values are larger than, albeit close to, the ideal value for a cubic perovskite structure ($r = 1$), and for such values cubic perovskites are the expectation\cite{Vasala2015}. 

\begin{figure}[ht!]
\includegraphics[scale = 0.60]{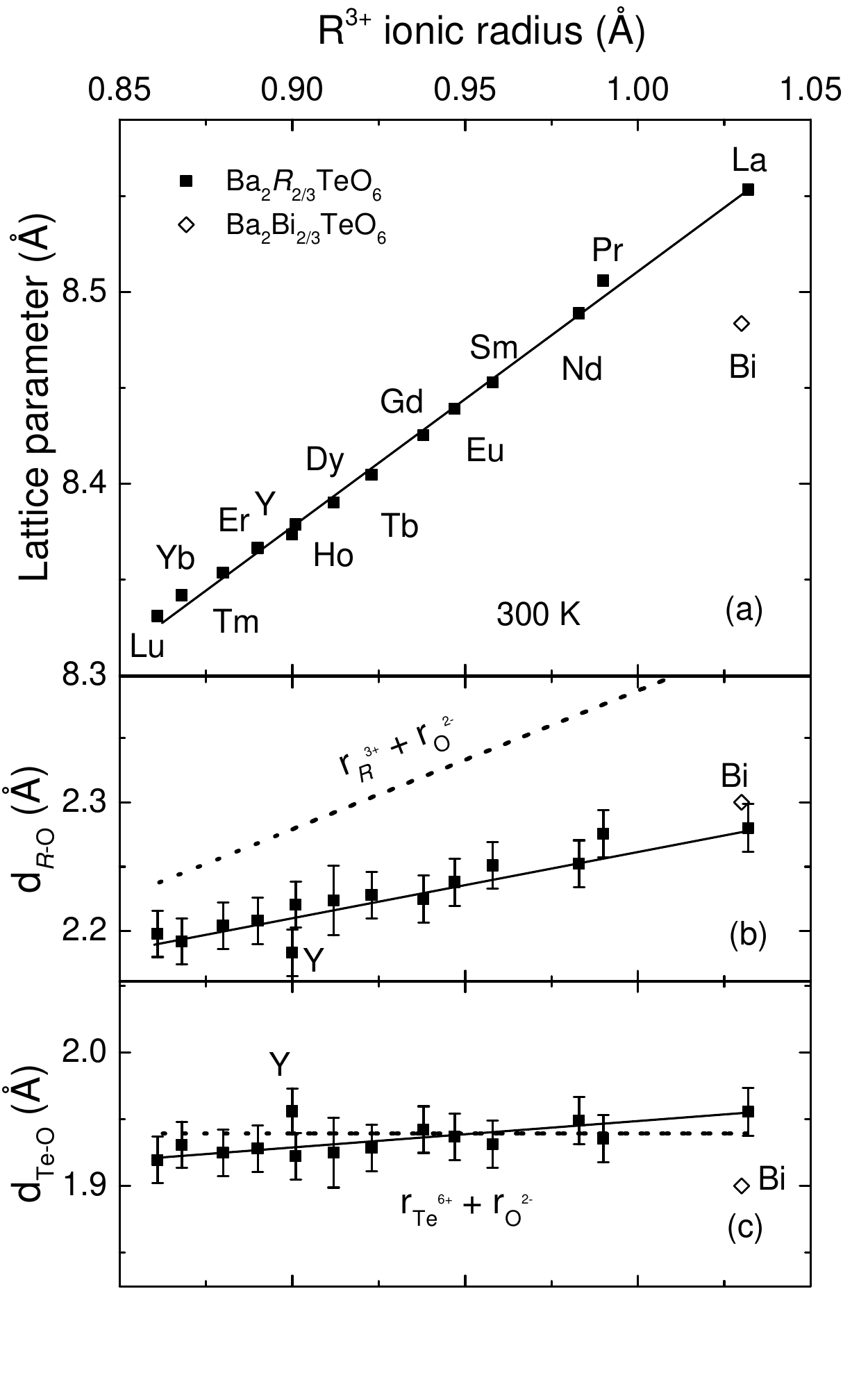}
\caption{(a) Lattice parameters for Ba$_2R_{2/3}$TeO$_6$, (b) $R$-O bond lengths and (c) Te-O bond lengths as a function of ionic radius of $R^{3+}$ in a 6 coordinated polyhedron\cite{Shannon76}. Solid lines are guides to the eyes and dashed lines are expected values based on ionic radii\cite{Shannon76}. The lattice parameter and bond lengths of Ba$_2$Bi$_{2/3}$TeO$_6$\cite{Park00} with respect to the ionic radius of Bi$^{3+}$ is shown by a hollow diamond.}
\label{a}
\end{figure}

Similar Rietveld refinements were performed for all 14 compounds (all diffraction data are shown in the Supplementary Information).The refined lattice parameters, cell volume, oxygen coordinates, Te-O and $R$-O bond lengths are listed in Table~\ref{lattice}. The data are summarized in Fig.~\ref{a}. The figure shows the lattice parameters for Ba$_2R_{2/3}$TeO$_6$ plotted as a function of the Shannon and Prewitt (S$\&$P) $R^{3+}$ ionic radii\cite{Shannon1969,Shannon76}, as well as the actual measured $R$-O and Te-O separations. The ideal values for the latter two quantities, with the (S$\&$P) ionic radii for $R$O$_6$-coordination and r$_{O^{2-}}$ = 1.40 $\AA$ used for the calculation are also plotted, as is a comparison of these structural parameters to those found for Ba$_2$Bi$_{2/3}$TeO$_6$\cite{Park00}.The lattice parameters show a monotonic, essentially linear increase with increasing $R^{3+}$ ionic radius in this series, a trend that is consistent with the lanthanide contraction\cite{Taylor72}. The lattice parameter of Ba$_2$Y$_{2/3}$TeO$_6$ is similar to that of Ba$_2$Ho$_{2/3}$TeO$_6$, which is also consistent with their ionic radius. At the tabulated ionic radius for Bi$^{3+}$, however, the lattice parameter for Ba$_2$Bi$_{2/3}$TeO$_6$\cite{Park00} falls at a lower value than is found for the rare earth series. The Y-O and Te-O bond lengths obtained are anomalous-the former being lower than expected when compared to the other rare earths and the latter higher than expected. This is not clearly associated with a problem determining the oxygen positional parameter in this compound as the error  bars are no larger in this case than in the others. An interesting comparison, presenting a curious puzzle, is seen on comparison of the observed $R$-O and Te-O bond lengths vs. those calculated from the (S$\&$P) $R^{3+}$ radii for octahedral coordination plus the radius of O$^{2+}$ = 1.4 $\AA$. It can be seen that the $R$-O bond length variation is smaller than expected, and, while the Te-O bond length variation is approximately 1/3 that observed for the $R$-O bond length, and in the same direction, the values are very close to that expected from the (S$\&$P) radii. Naively, one would interpret the data to say that the Te-O bond is being stretched slightly by the rare earth octahedra. But that does not explain why the $R$-O bond lengths increase more slowly with the rare earth radii than expected. This discrepancy also cannot reasonably be accommodated by a relaxation of the oxygens towards the empty octahedral sites, as that relaxation should be constant across the $R$ series, determined by the O-O repulsion, and not a function of the $R^{3+}$ size. (As a constant value it cannot impact the slope of the line). Finally, the issue also cannot be resolved if the $R^{3+}$ radii in the B site for oxide perovskites is less than is seen more generally for 6-coordination- the slope of the line would still not match observations. Thus our observations suggest that an additional factor is in play, of potential interest for future study.

\begin{table}[!h]
\caption{Refined lattice parameters (a), unit cell volumes (V), oxygen atomic position ($x_O$), Te-O bond distances (d$_{Te-O}$) and $R$-O bond distances (d$_{R-O}$) of Ba$_2R_{2/3}$TeO$_6$ from x-ray powder diffraction data and measured effective moments.}
\begin{tabular}{p{0.7cm} p{1.4cm} p{1.0cm} p{1.2cm} p{1.3cm} p{1.3cm} p{1.3cm}}
\hline
$R^{3+}$ & a ($\AA$) & V ($\AA^3$)& $x_O$ & d$_{Te-O}$ ($\AA$) & d$_{R-O}$ ($\AA$) & $\mu_{eff}$ ($\mu_B$)\\
\hline
Y  & 8.3735(3) & 587.1 & 0.264(2) & 1.98(2) & 2.21(2) &  \\
La & 8.5533(3) & 625.8 & 0.269(2) & 1.98(2) & 2.30(2) & \\
Pr & 8.5060(3) & 615.4 & 0.270(2) & 1.96(2) & 2.30(2) & 3.7\\
Nd & 8.4889(3) & 611.7 & 0.268(2) & 1.97(2) & 2.27(2) & 3.3\\
Sm & 8.4528(3) & 604.0 & 0.269(2) & 1.95(2) & 2.27(2) & \\
Eu & 8.4389(3) & 601.0 & 0.268(2) & 1.96(2) & 2.26(2) & \\
Gd & 8.4253(3) & 598.1 & 0.267(2) & 1.96(2) & 2.25(2) & 7.8\\
Tb & 8.4047(4) & 593.7 & 0.268(2) & 1.95(2) & 2.25(2) & 9.5\\
Dy & 8.3903(4) & 590.6 & 0.268(3) & 1.95(3) & 2.25(3) & 10.4\\
Ho & 8.3787(4) & 588.2 & 0.268(2) & 1.94(2) & 2.24(2) & 10.5\\
Er & 8.3664(5) & 585.6 & 0.267(2) & 1.95(2) & 2.23(2) & 9.5\\
Tm & 8.3535(5) & 582.9 & 0.267(2) & 1.95(2) & 2.23(2) & 7.6\\
Yb & 8.3419(5) & 580.5 & 0.266(2) & 1.95(2) & 2.22(2) & 4.7\\
Lu & 8.3310(4) & 578.2 & 0.267(2) & 1.94(2) & 2.22(2) & \\
\hline
\end{tabular}
\label{lattice}
\end{table}

\begin{figure}[tbh!]
\includegraphics[scale = 0.35]{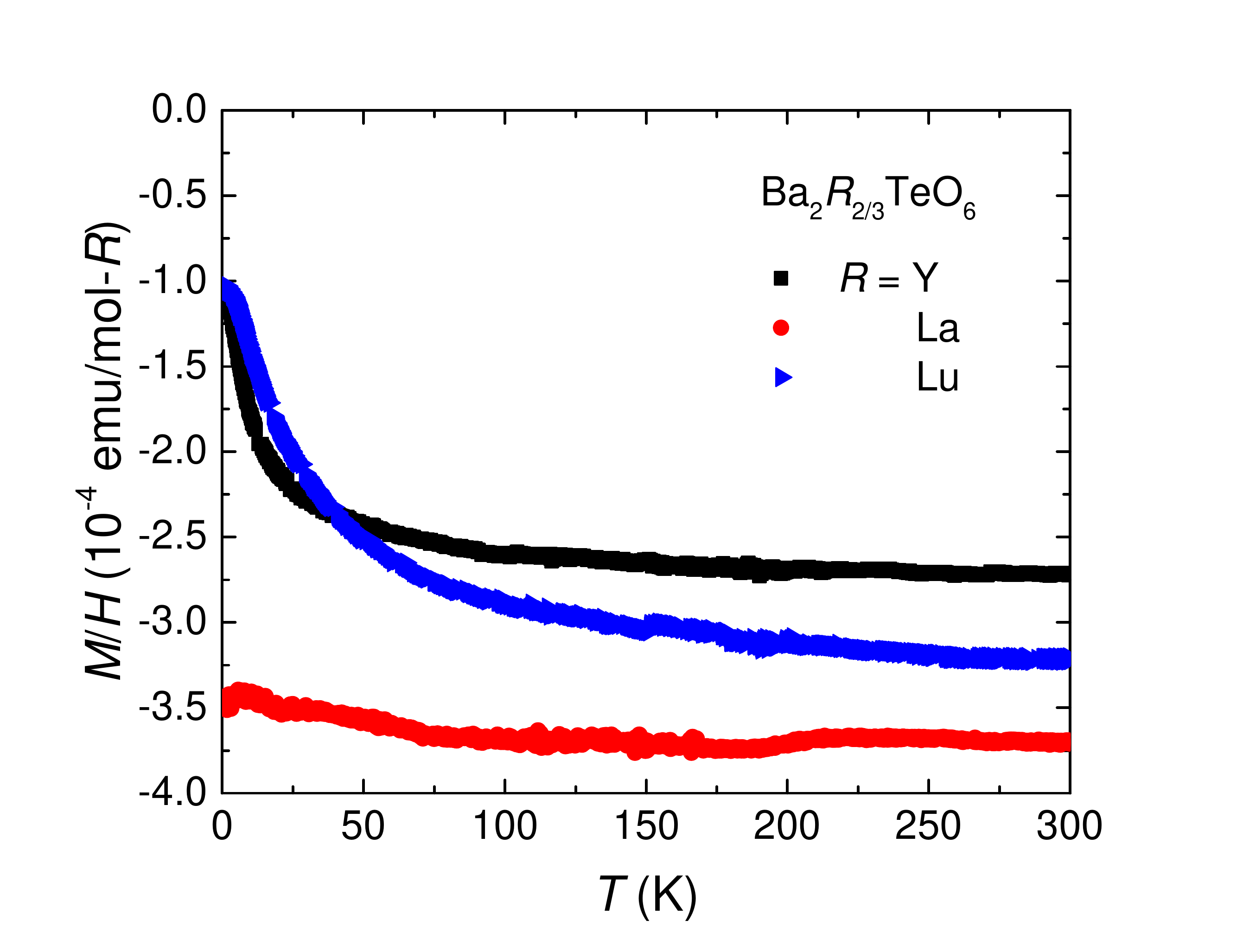}
\caption{(Color online) Temperature-dependent magnetic susceptibility of Ba$_2R_{2/3}$TeO$_6$ ($R$ = Y, La and Lu). The applied magnetic fields were 30 kOe, 50 kOe and 90 kOe for Y, La and Lu respectively.}
\label{YLaLu}
\end{figure}

Next, we consider the magnetic properties of Ba$_2R_{2/3}$TeO$_6$. In Fig.~\ref{YLaLu}, the temperature-dependent magnetic susceptibility of Ba$_2$Y$_{2/3}$TeO$_6$, Ba$_2$La$_{2/3}$TeO$_6$ and Ba$_2$Lu$_{2/3}$TeO$_6$ are shown. Because Y$^{3+}$, La$^{3+}$ and Lu$^{3+}$ bear no local moment due to a lack of un-paired 4f electrons, the magnetization of these three members are essentially temperature-independent. All of them are diamagnetic, primarily contributed by core diamagnetism, with a magnitude of $\sim$10$^{-4}$emu/mol-$R$. The magnetization upturn at low temperature for Ba$_2$Y$_{2/3}$TeO$_6$ and Ba$_2$Lu$_{2/3}$TeO$_6$ can be attributed to a very small amount paramagnetic impurity. For example, mixing 0.05$\%$ molar impurity of Dy into Lu would give rise to a similar Curie-Weiss like upturn in magnetization at low-temperature as shown in Fig.~\ref{YLaLu}.

\begin{figure}[tbh!]
\includegraphics[scale = 0.35]{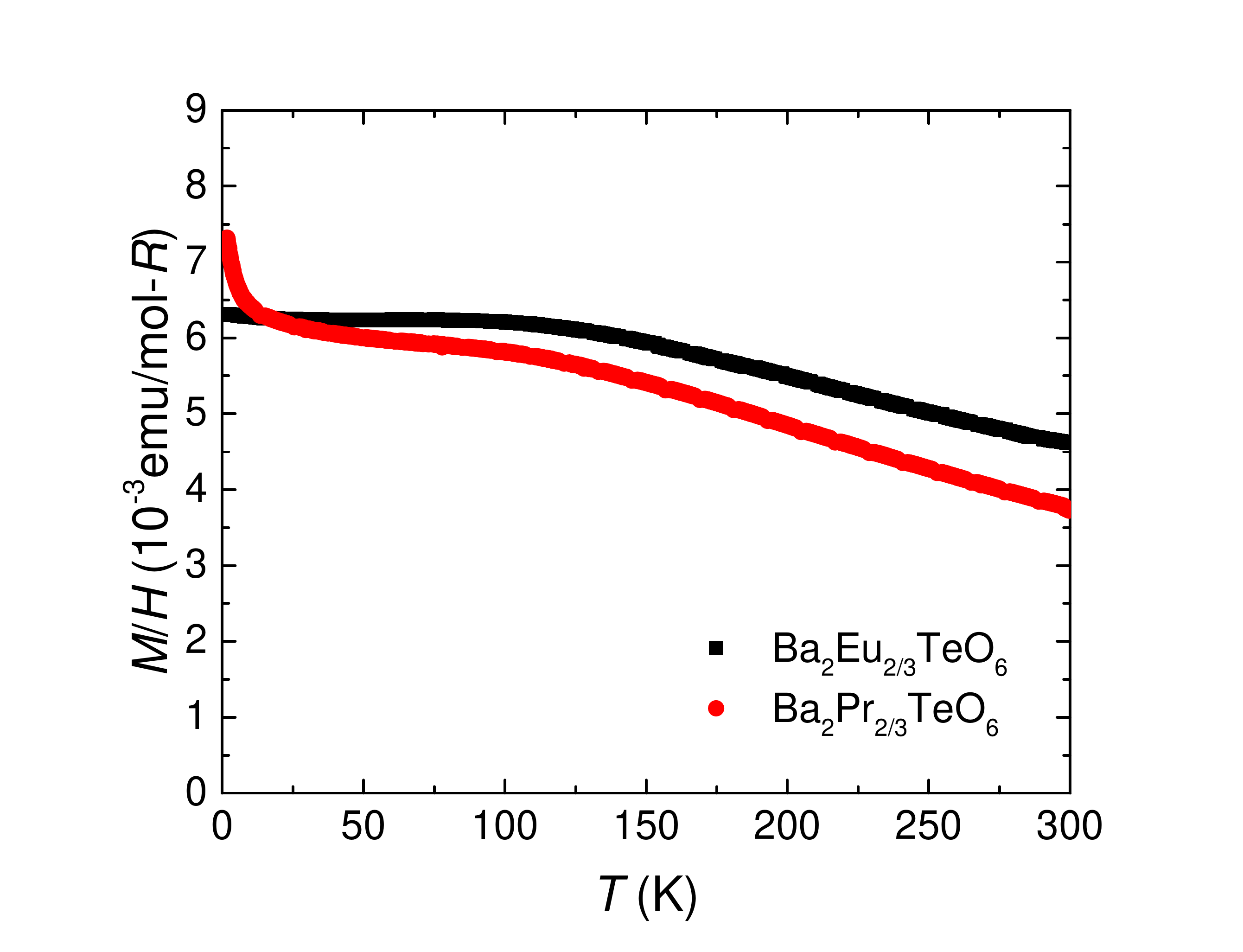}
\caption{(Color online) Temperature-dependent magnetic susceptibility of Ba$_2$Pr$_{2/3}$TeO$_6$, measured at 10 kOe and Ba$_2$Eu$_{2/3}$TeO$_6$, measured at 50 kOe.}
\label{EuM}
\end{figure}

Fig.~\ref{EuM} shows the temperature-dependent magnetic susceptibility of Ba$_2$Pr$_{2/3}$TeO$_6$. Above 250 K, the inverse magnetic susceptibility increases roughly linearly with temperature. A linear fit above 250 K based on Curie-Weiss law yields an effective moment of 3.7 $\mu_B$, which is close to the expected value (3.6 $\mu_B$). Deviation from linear behavior at lower temperatures may due to a large CEF splitting. No signature that can be associated with magnetic ordering was observed down to 1.8 K.

\begin{figure}[tbh!]
\includegraphics[scale = 0.35]{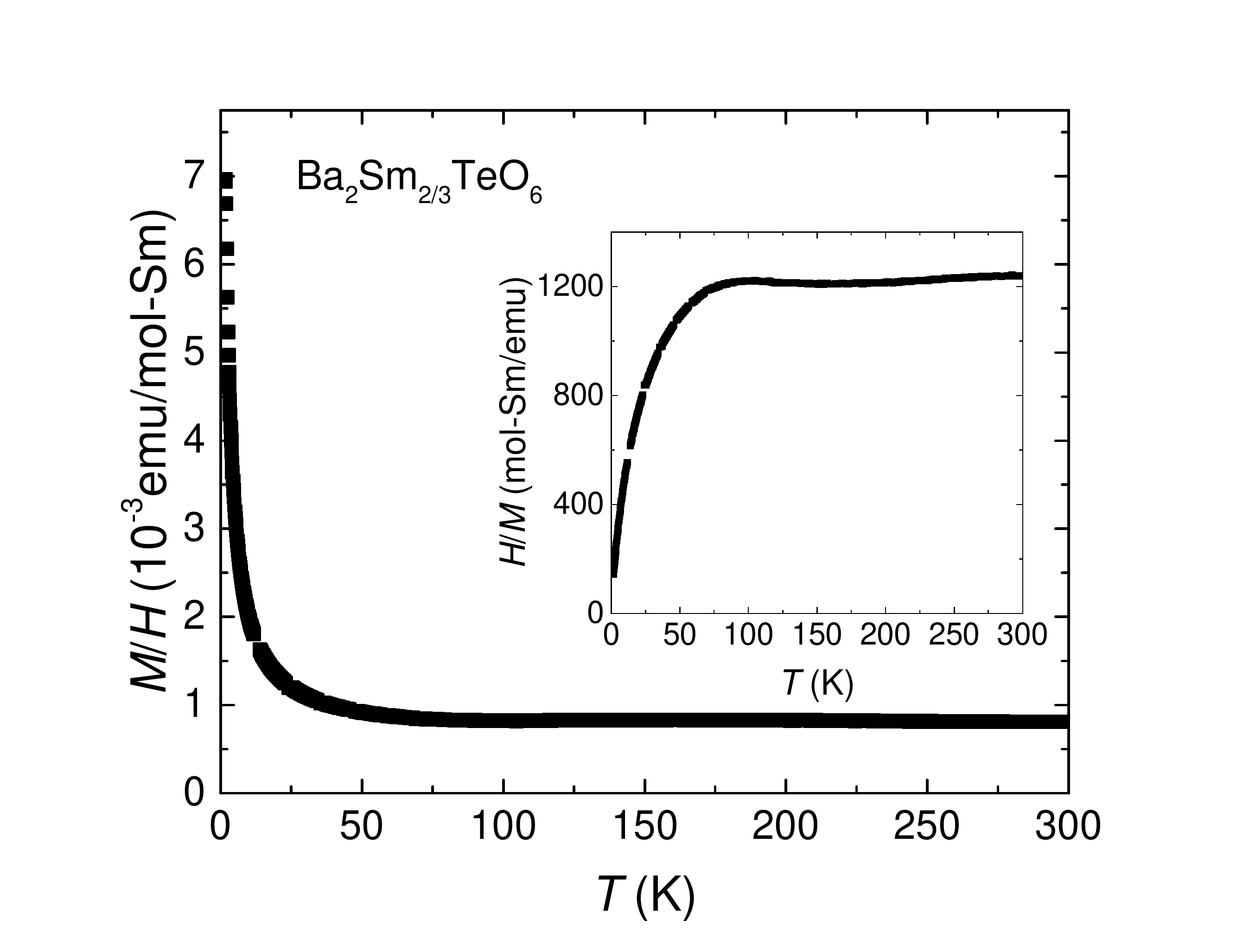}
\caption{Temperature-dependent magnetic susceptibility of Ba$_2$Sm$_{2/3}$TeO$_6$ measured at 90 kOe. Inset shows the temperature-dependent inverse magnetic susceptibility of Ba$_2$Sm$_{2/3}$TeO$_6$.}
\label{SmM}
\end{figure}

Temperature-dependent magnetic susceptibility of Ba$_2$Sm$_{2/3}$TeO$_6$ is shown in Fig.~\ref{SmM}. It shows a strong temperature-dependence. However, as manifested in the inset of Fig.~\ref{SmM}, the inverse magnetic susceptibility of Ba$_2$Sm$_{2/3}$TeO$_6$ does not exhibit a linear behavior in the high-temperature range. This was also seen in many other Sm based oxides\cite{Sanders16a,Sanders16b,Sanders17}, and is often attributed to a large contribution from Van Vleck magnetic susceptibility as the excited states of Sm$^{3+}$ are close to its ground state multiplets\cite{Taylor72}.

\begin{figure}[ht!]
\includegraphics[scale = 0.52]{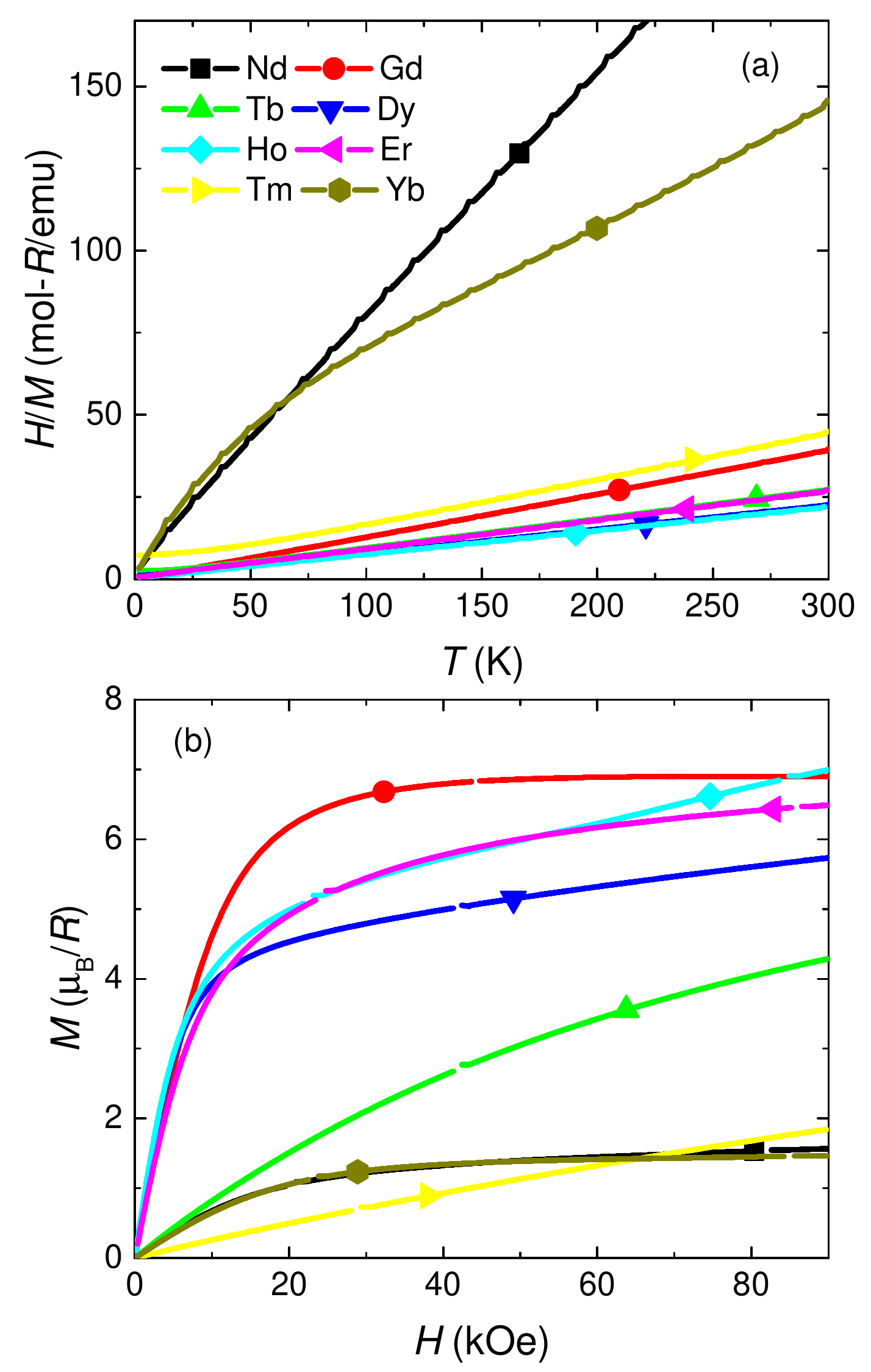}
\caption{(Color online) (a) Temperature-dependent inverse magnetic susceptibility and (b) field-dependent magnetization of Ba$_2R_{2/3}$TeO$_6$ ($R$ = Nd, Gd-Yb). The applied magnetic field shown in (a) were 20 kOe for $R$ =  Nd, Ho-Yb, and 30 kOe for $R$ = Gd-Dy. Data shown in (b) were measured at 1.8 K.}
\label{allMHT}
\end{figure}

The ground state of Eu$^{3+}$ does not bear a local moment because it has a total angular momentum $J$ = 0. The temperature-dependent magnetic susceptibility, shown in Fig.~\ref{EuM}, increases monotonically with decreasing temperature down to $\sim$100 K and then remains close to constant at lower temperatures. Similar to Sm$^{3+}$, the excited states of Eu$^{3+}$ are close to its non-magnetic ground state\cite{Taylor72}. Van Vleck paramagnetism, which is inversely proportional to the energy differences between states, thus becomes important for Sm$^{3+}$ and Eu$^{3+}$.

The other members of Ba$_2R_{2/3}$TeO$_6$ series all display a clear Curie-Weiss behavior at high-temperatures. The temperature-dependent inverse magnetic susceptibilities are shown in Fig.~\ref{allMHT}(a) and field-dependent magnetization measured at 1.8 K are shown in Fig.~\ref{allMHT}(b). The effective moments, obtained from high-temperature linear fits, are 3.3 $\mu_B$ for Ba$_2$Nd$_{2/3}$TeO$_6$, 7.8 $\mu_B$ for Ba$_2$Gd$_{2/3}$TeO$_6$, 9.5 $\mu_B$ for Ba$_2$Tb$_{2/3}$TeO$_6$, 10.4 $\mu_B$ for Ba$_2$Dy$_{2/3}$TeO$_6$, 10.5 $\mu_B$ for Ba$_2$Ho$_{2/3}$TeO$_6$, 9.5 $\mu_B$ for Ba$_2$Er$_{2/3}$TeO$_6$, 7.6 $\mu_B$ for Ba$_2$Tm$_{2/3}$TeO$_6$ and 4.7 $\mu_B$ for Ba$_2$Yb$_{2/3}$TeO$_6$. These values are all close to the expected values for $R^{3+}$. The similarity between experimental results and expectations supports the structural studies and simple charge balance concepts that indicate that the rare earth sites in Ba$_2R_{2/3}$TeO$_6$ are only 2/3 occupied.

Gd$^{3+}$ is magnetically isotropic due to a zero orbital angular momentum: $L$ = 0. Its temperature-dependent inverse magnetic susceptibility decreases linearly down to 1.8 K, showing no deviation due to CEF splitting. It also extrapolates to a Curie-Weiss temperature that is close to zero, indicating a very small magnetic interaction between Gd$^{3+}$. This weak interaction between rare earth ions can be understood from the structure point of view. Given the fact that un-paired 4$f$ electrons in $R^{3+}$ are spatially localized and have little direct overlap with oxygen $p$-orbitals, the interaction between rare earth ions lacks a viable channel in an insulator. In addition, each $R$O$_6$ octahedron is surrounded by six TeO$_6$ octahedra (with a full $d$ shell for Te$^{6+}$). Thus the TeO$_6$ octahedra simply increase the distance between neighboring rare earth ions, further weakening the $R-R$ magnetic dipole interaction. 

The field-dependent magnetization of Ba$_2$Gd$_{2/3}$TeO$_6$ reaches its saturation moment, 7 $\mu_B$, at 90 kOe. The other members shown in Fig.~\ref{allMHT} all exhibit various degrees of CEF depopulation as temperature decreases. This is illustrated by a deviation from linear inverse magnetic susceptibility at low temperatures. It is also illustrated in Fig.~\ref{allMHT}(b) where saturation moment is only reached for Ba$_2$Gd$_{2/3}$TeO$_6$. None of the Ba$_2R_{2/3}$TeO$_6$ shows signatures that can be associated with long range magnetic ordering or a short range spin glass state down to 1.8 K.

\section{Conclusions}

In conclusion, we report the synthesis, structural and magnetic properties of a new series of cubic double perovskite: Ba$_2R_{2/3}$TeO$_6$ ($R$ = Y, La, Pr, Nd, Sm-Lu). All compounds were synthesized via solid state reaction in air. Structurally, trivalent rare earth ions and Te$^{6+}$ undertake a standard alternating ordering on the B sites of the double perovskite structure-the K$_2$NaAlF$_6$ structure type. The rare earth ion sites in Ba$_2R_{2/3}$TeO$_6$ are only 2/3 occupied to balance the charge. This occupancy is also confirmed by temperature-dependent magnetization measurements where the measured magnetic effective moments are close to theoretically calculated values. Lattice parameters for Ba$_2R_{2/3}$TeO$_6$ decrease linearly with decreasing rare earth ionic radius, following the lanthanide contraction. Further research may be of interest to determine the reason for the apparent discrepancy between the $R$-O distances found here and the expected based on the S$\&$P tables. None of the discovered compounds order magnetically down to 1.8 K.

\section{Acknowledgements}

We would like to thank W.W. Xie and E. Carnicom for useful discussions. This work was supported by the Gordon and Betty Moore EPiQS initiative, grant number GBMF-4412.

\bibliographystyle{apsrev4-1}

\begin{thebibliography}{19}%
\makeatletter
\providecommand \@ifxundefined [1]{%
 \@ifx{#1\undefined}
}%
\providecommand \@ifnum [1]{%
 \ifnum #1\expandafter \@firstoftwo
 \else \expandafter \@secondoftwo
 \fi
}%
\providecommand \@ifx [1]{%
 \ifx #1\expandafter \@firstoftwo
 \else \expandafter \@secondoftwo
 \fi
}%
\providecommand \natexlab [1]{#1}%
\providecommand \enquote  [1]{``#1''}%
\providecommand \bibnamefont  [1]{#1}%
\providecommand \bibfnamefont [1]{#1}%
\providecommand \citenamefont [1]{#1}%
\providecommand \href@noop [0]{\@secondoftwo}%
\providecommand \href [0]{\begingroup \@sanitize@url \@href}%
\providecommand \@href[1]{\@@startlink{#1}\@@href}%
\providecommand \@@href[1]{\endgroup#1\@@endlink}%
\providecommand \@sanitize@url [0]{\catcode `\\12\catcode `\$12\catcode
  `\&12\catcode `\#12\catcode `\^12\catcode `\_12\catcode `\%12\relax}%
\providecommand \@@startlink[1]{}%
\providecommand \@@endlink[0]{}%
\providecommand \url  [0]{\begingroup\@sanitize@url \@url }%
\providecommand \@url [1]{\endgroup\@href {#1}{\urlprefix }}%
\providecommand \urlprefix  [0]{URL }%
\providecommand \Eprint [0]{\href }%
\providecommand \doibase [0]{http://dx.doi.org/}%
\providecommand \selectlanguage [0]{\@gobble}%
\providecommand \bibinfo  [0]{\@secondoftwo}%
\providecommand \bibfield  [0]{\@secondoftwo}%
\providecommand \translation [1]{[#1]}%
\providecommand \BibitemOpen [0]{}%
\providecommand \bibitemStop [0]{}%
\providecommand \bibitemNoStop [0]{.\EOS\space}%
\providecommand \EOS [0]{\spacefactor3000\relax}%
\providecommand \BibitemShut  [1]{\csname bibitem#1\endcsname}%
\let\auto@bib@innerbib\@empty
\bibitem [{\citenamefont {King}\ and\ \citenamefont {Woodward}(2010)}]{King10}%
  \BibitemOpen
  \bibfield  {author} {\bibinfo {author} {\bibfnamefont {G.}~\bibnamefont
  {King}}\ and\ \bibinfo {author} {\bibfnamefont {P.~M.}\ \bibnamefont
  {Woodward}},\ }\href {\doibase 10.1039/B926757C} {\bibfield  {journal}
  {\bibinfo  {journal} {J. Mater. Chem.}\ }\textbf {\bibinfo {volume} {20}},\
  \bibinfo {pages} {5785} (\bibinfo {year} {2010})}\BibitemShut {NoStop}%
\bibitem [{\citenamefont {Bednorz}\ and\ \citenamefont
  {M{\"u}ller}(1986)}]{Bednorz1986}%
  \BibitemOpen
  \bibfield  {author} {\bibinfo {author} {\bibfnamefont {J.~G.}\ \bibnamefont
  {Bednorz}}\ and\ \bibinfo {author} {\bibfnamefont {K.~A.}\ \bibnamefont
  {M{\"u}ller}},\ }\href {\doibase 10.1007/BF01303701} {\bibfield  {journal}
  {\bibinfo  {journal} {Z. Phys. B}\ }\textbf {\bibinfo {volume} {64}},\
  \bibinfo {pages} {189} (\bibinfo {year} {1986})}\BibitemShut {NoStop}%
\bibitem [{\citenamefont {Cava}(2000)}]{Cava00}%
  \BibitemOpen
  \bibfield  {author} {\bibinfo {author} {\bibfnamefont {R.~J.}\ \bibnamefont
  {Cava}},\ }\href {\doibase 10.1111/j.1151-2916.2000.tb01142.x} {\bibfield
  {journal} {\bibinfo  {journal} {J. Am. Ceram. Soc.,}\ }\textbf {\bibinfo
  {volume} {83}},\ \bibinfo {pages} {5} (\bibinfo {year} {2000})}\BibitemShut
  {NoStop}%
\bibitem [{\citenamefont {Kobayashi}\ \emph {et~al.}(1998)\citenamefont
  {Kobayashi}, \citenamefont {Kimura}, \citenamefont {Sawada}, \citenamefont
  {Terakura},\ and\ \citenamefont {Tokura}}]{Kobayashi98}%
  \BibitemOpen
  \bibfield  {author} {\bibinfo {author} {\bibfnamefont {K.-I.}\ \bibnamefont
  {Kobayashi}}, \bibinfo {author} {\bibfnamefont {T.}~\bibnamefont {Kimura}},
  \bibinfo {author} {\bibfnamefont {H.}~\bibnamefont {Sawada}}, \bibinfo
  {author} {\bibfnamefont {K.}~\bibnamefont {Terakura}}, \ and\ \bibinfo
  {author} {\bibfnamefont {Y.}~\bibnamefont {Tokura}},\ }\href
  {http://dx.doi.org/10.1038/27167} {\bibfield  {journal} {\bibinfo  {journal}
  {Nature}\ }\textbf {\bibinfo {volume} {395}},\ \bibinfo {pages} {677}
  (\bibinfo {year} {1998})}\BibitemShut {NoStop}%
\bibitem [{\citenamefont {von Hippel}(1950)}]{vonHippel50}%
  \BibitemOpen
  \bibfield  {author} {\bibinfo {author} {\bibfnamefont {A.}~\bibnamefont {von
  Hippel}},\ }\href {\doibase 10.1103/RevModPhys.22.221} {\bibfield  {journal}
  {\bibinfo  {journal} {Rev. Mod. Phys.}\ }\textbf {\bibinfo {volume} {22}},\
  \bibinfo {pages} {221} (\bibinfo {year} {1950})}\BibitemShut {NoStop}%
\bibitem [{\citenamefont {Vasala}\ and\ \citenamefont
  {Karppinen}(2015)}]{Vasala2015}%
  \BibitemOpen
  \bibfield  {author} {\bibinfo {author} {\bibfnamefont {S.}~\bibnamefont
  {Vasala}}\ and\ \bibinfo {author} {\bibfnamefont {M.}~\bibnamefont
  {Karppinen}},\ }\href {\doibase
  https://doi.org/10.1016/j.progsolidstchem.2014.08.001} {\bibfield  {journal}
  {\bibinfo  {journal} {Prog. Solid St. Chem.}\ }\textbf {\bibinfo {volume}
  {43}},\ \bibinfo {pages} {1 } (\bibinfo {year} {2015})}\BibitemShut {NoStop}%
\bibitem [{\citenamefont {Park}\ and\ \citenamefont {Woodward}(2000)}]{Park00}%
  \BibitemOpen
  \bibfield  {author} {\bibinfo {author} {\bibfnamefont {J.-H.}\ \bibnamefont
  {Park}}\ and\ \bibinfo {author} {\bibfnamefont {P.~M.}\ \bibnamefont
  {Woodward}},\ }\href {\doibase
  http://dx.doi.org/10.1016/S1466-6049(99)00071-9} {\bibfield  {journal}
  {\bibinfo  {journal} {Int. J. Inorg. Mater.}\ }\textbf {\bibinfo {volume}
  {2}},\ \bibinfo {pages} {153 } (\bibinfo {year} {2000})}\BibitemShut
  {NoStop}%
\bibitem [{\citenamefont {Shannon}(1976)}]{Shannon76}%
  \BibitemOpen
  \bibfield  {author} {\bibinfo {author} {\bibfnamefont {R.~D.}\ \bibnamefont
  {Shannon}},\ }\href {\doibase 10.1107/S0567739476001551} {\bibfield
  {journal} {\bibinfo  {journal} {Acta Cryst.}\ }\textbf {\bibinfo {volume}
  {A32}},\ \bibinfo {pages} {751} (\bibinfo {year} {1976})}\BibitemShut
  {NoStop}%
\bibitem [{\citenamefont {Larson}\ and\ \citenamefont {Dreele}(2000)}]{GSAS}%
  \BibitemOpen
  \bibfield  {author} {\bibinfo {author} {\bibfnamefont {A.~C.}\ \bibnamefont
  {Larson}}\ and\ \bibinfo {author} {\bibfnamefont {R.~B.~V.}\ \bibnamefont
  {Dreele}},\ }\href@noop {} {\emph {\bibinfo {title} {{General Structure
  Analysis System (GSAS)}}}},\ \bibinfo {type} {Tech. Rep.}\ (\bibinfo
  {institution} {Los Alamos National Laboratory Report LAUR 86-748},\ \bibinfo
  {year} {2000})\BibitemShut {NoStop}%
\bibitem [{\citenamefont {Toby}(2001)}]{Toby2001}%
  \BibitemOpen
  \bibfield  {author} {\bibinfo {author} {\bibfnamefont {B.~H.}\ \bibnamefont
  {Toby}},\ }\href {\doibase 10.1107/S0021889801002242} {\bibfield  {journal}
  {\bibinfo  {journal} {J. Appl. Crystallogr.}\ }\textbf {\bibinfo {volume}
  {34}},\ \bibinfo {pages} {210} (\bibinfo {year} {2001})}\BibitemShut
  {NoStop}%
\bibitem [{\citenamefont {Momma}\ and\ \citenamefont {Izumi}(2011)}]{Momma11}%
  \BibitemOpen
  \bibfield  {author} {\bibinfo {author} {\bibfnamefont {K.}~\bibnamefont
  {Momma}}\ and\ \bibinfo {author} {\bibfnamefont {F.}~\bibnamefont {Izumi}},\
  }\href {\doibase 10.1107/S0021889811038970} {\bibfield  {journal} {\bibinfo
  {journal} {J. Appl. Cryst.}\ }\textbf {\bibinfo {volume} {44}},\ \bibinfo
  {pages} {1272} (\bibinfo {year} {2011})}\BibitemShut {NoStop}%
\bibitem [{\citenamefont {Moras}(1974)}]{Moras74}%
  \BibitemOpen
  \bibfield  {author} {\bibinfo {author} {\bibfnamefont {L.~R.}\ \bibnamefont
  {Moras}},\ }\href {\doibase http://dx.doi.org/10.1016/0022-1902(74)80190-9}
  {\bibfield  {journal} {\bibinfo  {journal} {J. inorg, nucL Chem.}\ }\textbf
  {\bibinfo {volume} {36}},\ \bibinfo {pages} {3876 } (\bibinfo {year}
  {1974})}\BibitemShut {NoStop}%
\bibitem [{\citenamefont {Anderson}\ \emph {et~al.}(1993)\citenamefont
  {Anderson}, \citenamefont {Greenwood}, \citenamefont {Taylor},\ and\
  \citenamefont {Poeppelmeier}}]{Anderson93}%
  \BibitemOpen
  \bibfield  {author} {\bibinfo {author} {\bibfnamefont {M.~T.}\ \bibnamefont
  {Anderson}}, \bibinfo {author} {\bibfnamefont {K.~B.}\ \bibnamefont
  {Greenwood}}, \bibinfo {author} {\bibfnamefont {G.~A.}\ \bibnamefont
  {Taylor}}, \ and\ \bibinfo {author} {\bibfnamefont {K.~R.}\ \bibnamefont
  {Poeppelmeier}},\ }\href {\doibase
  http://dx.doi.org/10.1016/0079-6786(93)90004-B} {\bibfield  {journal}
  {\bibinfo  {journal} {Prog. Solid St. Chem.}\ }\textbf {\bibinfo {volume}
  {22}},\ \bibinfo {pages} {197 } (\bibinfo {year} {1993})}\BibitemShut
  {NoStop}%
\bibitem [{\citenamefont {Goldschmidt}(1926)}]{Goldschmidt26}%
  \BibitemOpen
  \bibfield  {author} {\bibinfo {author} {\bibfnamefont {V.~M.}\ \bibnamefont
  {Goldschmidt}},\ }\href {\doibase 10.1007/BF01507527} {\bibfield  {journal}
  {\bibinfo  {journal} {Naturwissenschaften}\ }\textbf {\bibinfo {volume}
  {14}},\ \bibinfo {pages} {477} (\bibinfo {year} {1926})}\BibitemShut
  {NoStop}%
\bibitem [{\citenamefont {Shannon}\ and\ \citenamefont
  {Prewitt}(1969)}]{Shannon1969}%
  \BibitemOpen
  \bibfield  {author} {\bibinfo {author} {\bibfnamefont {R.~D.}\ \bibnamefont
  {Shannon}}\ and\ \bibinfo {author} {\bibfnamefont {C.~T.}\ \bibnamefont
  {Prewitt}},\ }\href {\doibase 10.1107/S0567740869003220} {\bibfield
  {journal} {\bibinfo  {journal} {Acta. Cryst.}\ }\textbf {\bibinfo {volume}
  {B25}},\ \bibinfo {pages} {925} (\bibinfo {year} {1969})}\BibitemShut
  {NoStop}%
\bibitem [{\citenamefont {Taylor}\ and\ \citenamefont
  {Darby}(1972)}]{Taylor72}%
  \BibitemOpen
  \bibfield  {author} {\bibinfo {author} {\bibfnamefont {K.~N.~R.}\
  \bibnamefont {Taylor}}\ and\ \bibinfo {author} {\bibfnamefont {M.~I.}\
  \bibnamefont {Darby}},\ }\href@noop {} {\emph {\bibinfo {title} {{Physics of
  Rare Earth Solids}}}}\ (\bibinfo  {publisher} {Chapman and Hall Ltd,
  London},\ \bibinfo {year} {1972})\BibitemShut {NoStop}%
\bibitem [{\citenamefont {Sanders}\ \emph
  {et~al.}(2016{\natexlab{a}})\citenamefont {Sanders}, \citenamefont {Krizan},\
  and\ \citenamefont {Cava}}]{Sanders16a}%
  \BibitemOpen
  \bibfield  {author} {\bibinfo {author} {\bibfnamefont {M.~B.}\ \bibnamefont
  {Sanders}}, \bibinfo {author} {\bibfnamefont {J.~W.}\ \bibnamefont {Krizan}},
  \ and\ \bibinfo {author} {\bibfnamefont {R.~J.}\ \bibnamefont {Cava}},\
  }\href {\doibase 10.1039/C5TC03798K} {\bibfield  {journal} {\bibinfo
  {journal} {J. Mater. Chem. C}\ }\textbf {\bibinfo {volume} {4}},\ \bibinfo
  {pages} {541} (\bibinfo {year} {2016}{\natexlab{a}})}\BibitemShut {NoStop}%
\bibitem [{\citenamefont {Sanders}\ \emph
  {et~al.}(2016{\natexlab{b}})\citenamefont {Sanders}, \citenamefont {Baroudi},
  \citenamefont {Krizan}, \citenamefont {Mukadam},\ and\ \citenamefont
  {Cava}}]{Sanders16b}%
  \BibitemOpen
  \bibfield  {author} {\bibinfo {author} {\bibfnamefont {M.~B.}\ \bibnamefont
  {Sanders}}, \bibinfo {author} {\bibfnamefont {K.~M.}\ \bibnamefont
  {Baroudi}}, \bibinfo {author} {\bibfnamefont {J.~W.}\ \bibnamefont {Krizan}},
  \bibinfo {author} {\bibfnamefont {O.~A.}\ \bibnamefont {Mukadam}}, \ and\
  \bibinfo {author} {\bibfnamefont {R.~J.}\ \bibnamefont {Cava}},\ }\href
  {\doibase 10.1002/pssb.201600256} {\bibfield  {journal} {\bibinfo  {journal}
  {physica status solidi (b)}\ }\textbf {\bibinfo {volume} {253}},\ \bibinfo
  {pages} {2056} (\bibinfo {year} {2016}{\natexlab{b}})}\BibitemShut {NoStop}%
\bibitem [{\citenamefont {Sanders}\ \emph {et~al.}(2017)\citenamefont
  {Sanders}, \citenamefont {Cevallos},\ and\ \citenamefont {Cava}}]{Sanders17}%
  \BibitemOpen
  \bibfield  {author} {\bibinfo {author} {\bibfnamefont {M.~B.}\ \bibnamefont
  {Sanders}}, \bibinfo {author} {\bibfnamefont {F.~A.}\ \bibnamefont
  {Cevallos}}, \ and\ \bibinfo {author} {\bibfnamefont {R.~J.}\ \bibnamefont
  {Cava}},\ }\href {http://stacks.iop.org/2053-1591/4/i=3/a=036102} {\bibfield
  {journal} {\bibinfo  {journal} {Mater. Res. Express}\ }\textbf {\bibinfo
  {volume} {4}},\ \bibinfo {pages} {036102} (\bibinfo {year}
  {2017})}\BibitemShut {NoStop}%
\end{thebibliography}

%

\end{document}